\documentstyle[11pt]{article}
\begin{document}

\vspace*{0.5in}
{\Large
\centerline {\bf Quantum Wavelet Transforms: Fast Algorithms} 
\centerline {\bf and Complete Circuits\footnote{Presented at 1st NASA Int. Conf. on Quantum Computing and Communication, Palm Spring, CA, Feb. 17-21, 1998.}}               
}
\vspace{0.5in}

\centerline {\bf Amir Fijany and Colin P. Williams}
\vspace{0.1in}
\centerline {\it  Jet Propulsion Laboratory, California Institute of Technology}
\centerline {4800 Oak Grove Drive, Pasadena, CA 91109}
\centerline {Email: Amir.Fijany@jpl.nasa.gov \, and \, Colin.P.Williams@jpl.nasa.gov}          
\vspace{0.2in}
\centerline {\bf Abstract}

The quantum Fourier transform (QFT), a quantum analog of the classical Fourier transform, has been shown to be a powerful tool in developing quantum algorithms. However, in classical computing there is another class of unitary transforms, the wavelet transforms, which are every bit as useful as the Fourier transform.  Wavelet transforms are used to expose the
multi-scale structure of a signal and are likely to be useful for quantum image processing and quantum data compression. In this paper, we derive efficient, complete, quantum circuits for two representative quantum wavelet transforms, the quantum Haar and quantum Daubechies $D^{(4)}$ transforms. Our approach is to factor the classical operators for these transforms into direct sums, direct products and dot products of unitary matrices. In so doing, we find that permutation matrices, a particular class of unitary matrices, play a pivotal role.  Surprisingly, we find that operations that are easy and inexpensive to implement classically are not always easy and inexpensive to implement quantum mechanically, and vice versa. In particular, the computational cost of performing certain permutation matrices is ignored classically because they can be avoided explicitly.  However, quantum mechanically, these permutation operations must be performed explicitly and hence their cost enters into the full complexity measure of the quantum transform. We consider the particular set of permutation matrices arising in quantum wavelet transforms and develop efficient quantum circuits that implement them. This allows us to design efficient, complete quantum circuits for the quantum wavelet transform.
   
\vspace{0.1in}
 

\section{Introduction}

The field of quantum computing has undergone an explosion of activity over the past few years. Several important quantum algorithms are now known. Moreover, prototypical quantum computers have already been built using nuclear magnetic resonance [1, 2] and nonlinear optics technologies [3]. Such devices are far from being general-purpose computers.
Nevertheless, they constitute significant milestones along the road to practical quantum computing. 

A quantum computer is a physical device whose natural evolution over time can be interpreted as the execution of a useful computation. The basic element of a quantum computer is the quantum bit or "qubit", implemented physically as the state of some convenient 2-state quantum system such as the spin of an electron.  Whereas a classical bit must be either a 0 or a 1
at any instant, a qubit is allowed to be an arbitrary superposition of a 0 and a 1 simultaneously. To make a quantum memory register we simply consider the simultaneous state of (possibly entangled) tuples of qubits.

The state of a quantum memory register, or any other isolated quantum system, evolves in time according to some unitary operator. Hence, if the evolved state of a quantum memory register is interpreted as having implemented some computation, that computation must be describable as a unitary operator.  If the quantum memory register consists of $n$ qubits, this operator will be represented, mathematically, as some $2^n \times 2^n$ dimensional unitary matrix. 

Several quantum algorithms are now known, the most famous examples being Deutsch and Jozsa's algorithm for deciding whether a function is even or balanced [4], Shor's algorithm for factoring a composite integer [5] and Grover's algorithm for finding an item in an unstructured database [6]. However, the field is growing rapidly and new quantum algorithms are
being discovered every year.  Some recent examples include Brassard, Hoyer, and Tapp's quantum algorithm for counting the number of solutions to a problem [7], Cerf, Grover and Williams quantum algorithm for solving NP-complete problems by nesting one quantum search within another [8] and van Dam, Hoyer, and Tapp's algorithm for distributed quantum computing [9].

The fact that quantum algorithms are describable in terms of unitary transformations is both good news and bad for quantum computing. The good news is that knowing that a quantum computer must perform a unitary transformation allows theorems to be proved about the tasks that quantum computers can and cannot do. For example, Zalka has proved that Grover's algorithm is optimal [10]. Aharonov, Kitaev, and Nisan have proved that a quantum algorithm that involves intermediate measurements is no more powerful than one that postpones all measurements until the end of the unitary evolution stage [11].  Both these proofs rely upon quantum algorithms being unitary transformations. On the other hand, the bad news is that many computations that we would like to perform are not originally described in terms of unitary operators. For example, a desired computation might be nonlinear, irreversible or both nonlinear and irreversible. As a unitary transformation must be linear and reversible we might need to be quite creative in encoding a desired computation on a quantum computer. Irreversibility can be handled by incorporating extra "ancilla" qubits that permit us to remember the input corresponding to each output. But nonlinear transformations are still problematic.

Fortunately, there is an important class of computations, the unitary transforms, such as the Fourier transform, Walsh-Hadamard transform and assorted wavelet transforms, that are describable, naturally, in terms of unitary operators. Of these, the Fourier and Walsh-Hadamard transforms have been the ones studied most extensively by the quantum computing community. In fact, the quantum Fourier transform (QFT) is now recognized as being pivotal in many known quantum algorithms  [12]. The quantum Walsh-Hadamard transform is a critical component of both Shor's algorithm [5] and Grover's algorithm [6]. However, the wavelet transforms are every bit as useful as the Fourier transform, at least in the context of classical computing. For example, wavelet transforms are particularly suited to exposing the multi-scale structure of a signal. They are likely to be useful for quantum image processing and quantum data compression. It is natural therefore to consider how to achieve a quantum wavelet transform.

Starting with the unitary operator for the wavelet transform, the next step in the process of finding a quantum circuit that implements it, is to factor the wavelet operator into the direct sum, direct product and dot product of smaller unitary operators.  These operators correspond to 1-qubit and 2-qubit quantum gates. For such a circuit to be physically realizable, the number of gates within it must be bounded above by a polynomial in the number of qubits, $n$.  Finding such a factorization can be extremely challenging. For example, although there are known algebraic techniques for factoring an arbitrary $2^n \times 2^n$ operator, e.g. [13], they are guaranteed to produce $O(2^n)$, i.e., exponentially many, terms in the factorization. Hence, although such a factorization is mathematically valid, it is physically unrealizable because, when treated as a quantum circuit design, would require too many quantum gates. Indeed, Knill has {\it proved} that an arbitrary unitary matrix will require exponentially many quantum gates if we restrict ourselves to using only gates that correspond to all 1-qubit rotations and XOR [14]. It is therefore clear that the key enabling factor for achieving an efficient quantum implementation, i.e., with a polynomial time and space complexity, is to exploit the specific structure of the given unitary operator. 

Perhaps the most striking example of the potential for achieving compact and efficient quantum circuits is the case of the Walsh-Hadamard transform. In quantum computing, this transform arises whenever a quantum register is loaded with all integers in the range 0 to $2^n-1$. Classically, application of the Walsh-Hadamard transform on a vector of length $2^n$
involves a complexity of $O(2^n)$. Yet, by exploiting the factorization of the Walsh-Hadamard operator in terms of the Kroenecker product, it can implemented with a complexity of $O(1)$ by $n$ identical 1-qubit quantum gates. Likewise, the classical FFT algorithm has been found to be implementable in a polynomial space and time complexity, quantum circuit
[15] (see also Sec. 2.3). However, exploitation of the operator structure arising in the wavelet transforms (and perhaps other unitary transforms) is more challenging.

A key technique, in classical computing, for exposing and exploiting specific structure of a given unitary transform is the use of permutation matrices. In fact, there is an extensive literature in classical computing on the use of permutation matrices for factorizing unitary transforms into simpler forms that enable efficient implementations to be
devised (see, for example, [16] and [17]). However, the underlying assumption in using the permutation matrices in classical computation is that they can be implemented easily and inexpensively. Indeed, they are considered so trivial that the cost of their implementation is often not included in the complexity analysis. This is because any permutation matrix
can be described by its effect on the ordering of the elements of a vector. Hence, it can simply be implemented by re-ordering the elements of the vector involving only data movement and without performing any arithmetic operations. As is shown in this paper, the permutation matrices also play a pivotal role in the factorization of the unitary operators that arise in the wavelet transforms. However, unlike the classical computing, the cost of implementation of the permutation matrices cannot be neglected in quantum computing. Indeed, the main issue in deriving feasible and efficient quantum circuits for the quantum wavelet transforms considered in this paper, is the design of efficient quantum circuits for certain permutation matrices. Note that, any permutation matrix acting on $n$ qubits can mathematically be represented by a $2^n \times 2^n$ unitary operator. Hence, it is possible to factor any permutation matrix by using general techniques such as [13] but this would lead to an exponential time and space complexity. However, the permutation matrices, due to their specific structure (i.e., sparsity pattern), represents a very special subclass of unitary matrices. Therefore, the key to achieve an efficient quantum implementation of permutation matrices is the exploitation of this specific structure.

In this paper, we first develop efficient quantum circuits for a set of permutation matrices arising in the development of the quantum wavelet transforms (and the quantum Fourier transform). We propose three techniques for an efficient quantum implementation of permutation matrices, depending on the permutation matrix considered. In the first technique, we show that a certain class of permutation matrices, designated as {\it qubit permutation matrices}, can directly be described by their effect on the ordering of qubits. This quantum description is very similar to classical description of the permutation matrices. We show that the {\it Perfect Shuffle} permutation matrix, designated as $\Pi_{2^n}$, and the {\it Bit Reversal} permutation matrix, designated as $P_{2^n}$, which arise in the quantum wavelet and Fourier transforms (as well as in many other classical computations) belong to this class. We present a new gate, designated as the {\it qubit swap gate} or $\Pi_4$, which can be used to directly derive efficient quantum circuits for implementation of the qubit permutation matrices. Interestingly, such circuits for quantum implementation of $\Pi_{2^n}$ and $P_{2^n}$ lead to new factorizations of these two permutation matrices which were not previously know in classical computation. A second technique is based on a {\it quantum arithmetic description} of permutation matrices. In particular, we consider the {\it downshift} permutation matrix, designated as $Q_{2^n}$, which plays a major role in derivation of quantum wavelet transforms and also frequently arises in many classical computations [16]. We show that a quantum description of $Q_{2^n}$ can be given as a {\it quantum arithmetic operator}. This description then allows the quantum implementation of $Q_{2^n}$ by using the quantum arithmetic circuits proposed in [18]. 

A third technique is based on developing totally new factorizations of the permutation matrices. This technique is the most case dependent, challenging, and even counterintuitive (from a classical computing point of view). For this technique, we again consider the permutation matrix $Q_{2^n}$ and we show that it can be factored in terms of FFT which then allows its implementation by using the circuits for QFT. More interestingly, however, we derive a recursive factorization of $Q_{2^n}$ which was not previously known in classical computation. This new factorization enables a direct and efficient implementation of $Q_{2^n}$. Our analysis of though a limited set of permutation matrices reveals some of the surprises of quantum computing in contrast to classical computing. That is, certain operations that are hard to implement in classical computing are much easier to implement on quantum computing and vice versa. As a specific example, while the classical  implementation of $\Pi_{2^n}$ and $P_{2^n}$ are much harder (in terms of the data movement pattern) than $Q_{2^n}$, their quantum implementation is much easier and more straightforward than $Q_{2^n}$.

Given a wavelet kernel, its application is usually performed according to the packet or pyramid algorithms. Efficient quantum implementation of theses two algorithms requires efficient circuits for operators of the form $I_{2^{n-i}}  \otimes \Pi_{2^i}$ and $\Pi_{2^i} \oplus I_{2^n - 2^i}$, for some $i$, where $\otimes$ and $\oplus$ designate, respectively, the kronecker product and the direct sum operator. We show that these operators can be efficiently implemented by using our proposed circuits for implementation of $\Pi_{2^i}$. We then consider two representative wavelet kernels, the Haar [17] and Daubechies $D^{(4)}$ [19] wavelets which have previously been considered by Hoyer [20]. For the Haar wavelet, we show that Hoyer's proposed solution is incomplete since it does not lead to a gate-level circuit and, consequently, it does not allow the analysis of time and space complexity. We propose a scheme for design of a complete gate-level circuit for the Haar wavelet and analyze its time and space complexity. For the Daubechies $D^{(4)}$ wavelet, we develop three new factorizations   which lead to three gate-level circuits for its implementation. Interestingly, one of this factorization allows efficient implementation of Daubechies $D^{(4)}$ wavelets by using the circuit for QFT.                       
       
\clearpage

\section{Efficient Quantum Circuits for two Fundamental Qubits Permutation Matrices: Perfect Shuffle and Bit-Reversal}

In this section, we develop quantum circuits for two fundamental permutation matrices, the perfect shuffle, $\Pi_{2^n}$,  and the bit reversal, $P_{2^n}$, permutation matrices, which arise in quantum wavelet and Fourier transforms as well as many classical computations involving unitary transforms for signal and image processing [16]. For quantum computing, these two permutation matrices can directly be described in terms of their effect on ordering of qubits. This enables the design of efficient circuits for their implementation. Interestingly, these circuits lead to the discovery of new factorizations for these two permutation matrices.     
 
\subsection{Perfect Shuffle Permutation Matrices}

A classical description of $\Pi_{2^n}$ can be given by describing its effect on a given vector. If $Z$ is a $2^n$-dimensional vector, then the vector $Y = \Pi_{2^n}Z$ is obtained by splitting $Z$ in half and then shuffling the top and bottom halves of the deck. Alternatively, a description of the matrix $\Pi_{2^n}$, in terms of its elements $\Pi_{ij}$, for $i$
and $j = 0, 1, \cdots, 2^n-1$, can be given as   
\begin{equation}
\Pi_{ij} = \left\{ \begin{array}{ll}
1 & \mbox{  if $j = i/2$ and $i$ is even, or if $j = (i - 1)/2 +2^{n-1}$ and 
$i$ is odd} \\
0 & \mbox{  otherwise}
\end{array}
\right.
\end{equation}
As first noted by Hoyer [20], a quantum description of $\Pi_{2^n}$ can be given by    
\begin{equation}
\Pi_{2^n}: \, \vert a_{n-1} \, a_{n-2} \,  \cdots  \, a_1 \, a_0   \rangle \, 
\longmapsto \vert a_0 \, a_{n-1} \, a_{n-2} \, \cdots a_1   \rangle
\end{equation}
That is, for quantum computation, $\Pi_{2^n}$ is the operator which performs the left qubit-shift operation on $n$ qubits. Note that, $\Pi_{2^n}^t$ ($t$ indicates the transpose) performs the right qubit-shift operation, i.e.,
\begin{equation}
\Pi_{2^n}^t: \, \vert a_{n-1} \, a_{n-2} \, \cdots a_1 \, a_0 \rangle \, 
\longmapsto \vert a_{n-2} \, \cdots a_1 \, a_0 \, a_{n-1}   \rangle
\end{equation}

\subsection{Bit-Reversal Permutation Matrices}

A classical description of $P_{2^n}$ can be given by describing its effect on a given vector. If $Z$ is a $2^n$-dimensional vector and $Y = P_{2^n}Z$, then $Y_i = Z_j$, for $i = 0, 1, \cdots, 2^n-1$, wherein $j$ is obtained by reversing the bits in the binary representation of index $i$. Therefore, a description of the matrix $P_{2^n}$, in terms of its elements $P_{ij}$, for $i$ and $j = 0, 1, \cdots, 2^n-1$, is given as  
\begin{equation}
P_{ij} = \left\{ \begin{array}{cc} 
1 & \mbox{if $j$ is bit reversal of $i$} \\
0 & \mbox{otherwise}
\end{array}
\right.
\end{equation}
A factorization of $P_{2^n}$ in terms of $\Pi_{2^i}$ is given as [16] 
\begin{equation}
P_{2^n} = \Pi_{2^n}(I_2 \otimes \Pi_{2^{n-1}}) \cdots (I_{2^i} \otimes \Pi_{2^{n-i}}) \, \cdots (I_{2^{n-3}} \otimes \Pi_8)(I_{2^{n-2}} \otimes \Pi_4)
\end{equation}
A quantum description of $P_{2^n}$ is given as 
\begin{equation}
P_{2^n}: \, \vert a_{n-1} \, a_{n-2}, \cdots a_1 \, a_0  \rangle \, 
\longmapsto  \vert a_0 \, a_1 \, \cdots a_{n-2} \, a_{n-1} \rangle
\end{equation}
That is, $P_{2^n}$ is the operator which reverses the order of $n$ qubits. This quantum description can be seen from the factorization of $P_{2^n}$, given by (5), and quantum description of permutation matrices $\Pi_{2^i}$. It is interesting to note that for classical computation the term "bit-reversal" refers to reversing the bits in the binary representation of index of the elements of a vector while, for quantum computation, the matrix $P_{2^n}$ literally performs a reversal of the order of qubits.

Note that, $P_{2^n}$ is symmetric, i.e., $P_{2^n} = P_{2^n}^t$ [16]. This can be also easily proved based on the quantum description of $P_{2^n}$ since if the qubits are reversed twice then the original ordering of the qubits is restored. This implies that, $P_{2^n}P_{2^n} = I_{2^n}$ and since $P_{2^n}$ is orthogonal, i.e., $P_{2^n}P_{2^n}^t = I_{2^n}$, it then follows that $P_{2^n} = P_{2^n}^t$. 

\subsection{Quantum FFT and Bit-Reversal Permutation Matrix}

Here, we review the quantum FFT algorithm since it not only arises in derivation of the quantum wavelet transforms (see Sec. 4.3) but also it represents a case in which the roles of permutation matrices $\Pi_{2^n}$ and $P_{2^n}$ seems to have been overlooked in quantum computing literature.  

The classical Cooley-Tukey FFT factorization for a $2^n$-dimensional vector is given by [16] 
\begin{equation}
F_{2^n} = A_n A_{n-1} \cdots A_1 P_{2^n} = 
{\underline F}_{2^n} P_{2^n}
\end{equation}
where $A_i = I_{2^{n-i}} \otimes B_{2^i}$, $B_{2^i} = \frac {1} {\sqrt {2}}  
\left( \begin{array}{cc}
I_{2^{i-1}} & \Omega_{2^{i-1}} \\
I_{2^{i-1}} & - \Omega_{2^{i-1}} 
\end{array} \right)
$ and $\Omega_{2^{i-1}} = \mbox {Diag} \{1, \, \omega_{2^i}, \, \omega_{2^i}^2, \, \ldots , \omega_{2^i}^{2^{i-1} -1} \}$ with  
$\omega_{2^i} = e^{-{2 \iota \pi} \over {2^i}}$ and $\iota = \sqrt {-1}$. We have that $F_2 = W =
\frac {1} {\sqrt {2}}  
\left( \begin{array}{cc}
1 & 1 \\
1 & - 1 
\end{array} \right)
$. The operator 
\begin{equation}
{\underline F}_{2^n} = A_n A_{n-1} \cdots A_1 
\end{equation}
represents the computational kernel of Cooley-Tukey FFT while $P_{2^n}$ represents the permutation which needs to be performed on the elements of the input vector before feeding that vector into the computational kernel. Note that, the presence of $P_{2^n}$ in (7) is due to the accumulation of its factors, i.e., the terms $(I_{2^i} \otimes \Pi_{2^{n-i}})$, as given by (5).     

The Gentleman-Sande FFT factorization is obtained by exploiting the symmetry of  $F_{2^n}$ and transposing the Cooley-Tukey factorization [16] leading to
\begin{equation}
F_{2^n} = P_{2^n} A_1^t \cdots A_{n-1}^t A_n^t = 
P_{2^n} {\underline F}_{2^n}^t 
\end{equation}
where 
\begin{equation}
{\underline F}_{2^n}^t = A_1^t \cdots A_{n-1}^t A_n^t
\end{equation} 
represents the computational kernel of the Gentleman-Sande FFT while $P_{2^n}$ represents the permutation which needs to be performed to obtain the elements of the output vector in the correct order.

In [15] a quantum circuit for the implementation of ${\underline F}_{2^n}$, given by (8), is presented by developing a factorization of the operators $B_{2^i}$ as 
\begin{equation}
B_{2^i} =
\frac {1} {\sqrt {2}}  
\left( \begin{array}{cc}
I_{2^{i-1}} & \Omega_{2^{i-1}} \\
I_{2^{i-1}} & -\Omega_{2^{i-1}} 
\end{array} \right) = 
\frac {1} {\sqrt {2}}   
\left( \begin{array}{cc}
I_{2^{i-1}} & I_{2^{i-1}} \\
I_{2^{i-1}} & - I_{2^{i-1}} 
\end{array} \right)
\left( \begin{array}{cc}
I_{2^{i-1}} & 0 \\
0 & \Omega_{2^{i-1}} 
\end{array} \right)     
\end{equation}
Let $C_{2^i} = 
\left( \begin{array}{cc}
I_{2^{i-1}} & 0 \\
0 & \Omega_{2^{i-1}} 
\end{array} \right)
$. It then follows that 
\begin{equation}
B_{2^i} = (W \otimes I_{2^{i-1}}) C_{2^i} 
\end{equation}
\begin{equation}
A_i = I_{2^{n-i}} \otimes B_{2^i} = 
(I_{2^{n-i}} \otimes W \otimes I_{2^{i-1}})(I_{2^{n-i}} \otimes C_{2^i})
\end{equation}
In [15] a factorization of the operators $C_{2^i}$ is developed as 
\begin{equation}
C_{2^i} = \theta_{n-1, n-i}\theta_{n-2, n-i} \cdots \theta_{n-i+1, n-i}   
\end{equation}
where $\theta_{jk}$ is a two-bit gate acting on $j$th and $k$th qubits.   

Using (13)-(14) a circuit for implementation of (8) is developed in [15] and presented in Fig. 1. However, there is an error in the corresponding figure in [15] since it implies that, with a correct ordering of the input qubits, the output qubits are obtained in a reverse order. Note that, as can be seen from (7), the operator ${\underline F}_{2^n}$ performs the FFT operation and provides the output qubits in a correct order if the input qubits are presented in a reverse order.   

The quantum circuit for Gentleman-Sande FFT can be obtained from the circuit of Fig. 1 by first reversing the order of gates that build the operator block $A_i$ (and thus building operators $A_i^t$) and then reversing the order of the blocks representing operators $A_i$. By using the Gentleman-Sande circuit, with the input qubits in the correct order the output qubits are obtained in reverse order.

For an efficient and correct implementation of the quantum FFT, one needs to take into account the ordering of the input and output qubits, particularly if the FFT is used as a block box in a quantum computation. If the FFT is used as a stand-alone block or as the last stage in the computation (and hence its output is sampled directly), then it is more efficient to use the Gentleman-Sande FFT since the ordering of the output qubits does not cause any problem. If the FFT is used as the first stage of the computation, then it is more efficient to use the Cooley-Tukey factorization by preparing the input qubits in a reverse order. Note that, as in classical computation, each or a combination of the Cooley-Tukey or Gentleman-Sande FFT factorization can be chosen in a given quantum computation to avoid explicit implementation of $P_{2^n}$ (or, any other mechanism) for reversing the order of qubits and hence achieve a greater efficiency. As an example, in Sec. 4.3 we will show that the use of the Cooley-Tukey rather than the Gentleman-Sande factorization leads to a greater efficiency in quantum implementation by eliminating the need for an explicit implementation of $P_{2^n}$ (or, any other mechanism) for reversing the order of qubits.      

\subsection{A Basic Quantum Gate for Efficient Implementation of Qubits Permutation Matrices}

If a permutation matrix can be described by its effect on the ordering of the qubits then it might be possible to devise circuits for its implementation directly. We call the class of such permutation matrices as "Qubit Permutation Matrices". A set of efficient and practically realizable circuits for implementation of Qubit Permutation Matrices can be built by using a new quantum gate, called
{\it the qubit swap gate}, $\Pi_4$, where \begin{equation}
\Pi_4 = 
\left( \begin{array}{cccc}
1 & 0 & 0 & 0 \\
0 & 0 & 1 & 0 \\
0 & 1 & 0 & 0 \\
0 & 0 & 0 & 1 
\end{array} \right)  
\end{equation}
For quantum computation, $\Pi_4$ is the "qubit swap operator", i.e., 
\begin{equation}
\Pi_4: \,   \vert a_1 \, a_0 \rangle \, \longmapsto \, \vert a_0 \, a_1 \rangle
\end{equation}
The $\Pi_4$ gate, shown in Fig. 2.a, can be implemented with three XOR (or Controlled-NOT) gates as shown in Fig. 2.b. The  $\Pi_4$ gate offers two major advantages for practical implementation:
\begin{itemize}
\item It performs a local operation, i.e., swapping the two neighboring qubits. This locality can be advantageous in
practical realizations of quantum circuits, and  

\item Given the fact that $\Pi_4$ can be implemented using three XOR (or, Controlled-NOT) gates, it is possible to implement conditional operators involving $\Pi_4$, for example, operators of the form $\Pi_4 \oplus I_{2^n - 4}$, by using Controlled$^k$-NOT gates [21].   
\end{itemize}
   
A circuit for implementation of $\Pi_{2^n}$ by using $\Pi_4$ gates is shown in Fig. 3. This circuit is based on an   intuitively simple idea of successive swapping of the neighboring qubits, and implements $\Pi_{2^n}$ with a complexity of  $O(n)$ by using an $O(n)$ number of $\Pi_4$ gates. It is interesting to note that, this circuit leads to a new (to our knowledge) factorization of $\Pi_{2^n}$ in terms of $\Pi_4$ as 
\begin{equation}
\Pi_{2^n} = (I_{2^{n-2}} \otimes \Pi_4)(I_{2^{n-3}} \otimes \Pi_4 \otimes I_2)
\cdots (I_{2^{n-i}} \otimes \Pi_4 \otimes I_{2^{i-2}}) \, \cdots \, 
(I_2 \otimes \Pi_4 \otimes I_{2^{n-3}})(\Pi_4 \otimes I_{2^{n-2}})
\end{equation}
This new factorization of $\Pi_{2^n}$ is less efficient than other schemes (see, for example, [16]) for a {\it classical implementation} of $\Pi_{2^n}$. Interestingly, it is derived here as a result of our search for an efficient {\it quantum implementation} of $\Pi_{2^n}$, and in this sense it is only efficient for a quantum implementation. Note also, that a new (to our knowledge) recursive factorization of $\Pi_{2^i}$ directly results from Fig. (3) as  
\begin{equation}
\Pi_{2^i} = (I_{2^{i-2}} \otimes \Pi_4)(\Pi_{2^{i-1}} \otimes I_2)
\end{equation}

A circuit for implementation of $P_{2^n}$ by using $\Pi_4$ gates is shown in Fig. 4. Again, this circuit is based on an intuitively simple idea, that is, successive and parallel swapping of the neighboring qubits, and implements $P_{2^n}$ with a complexity of $O(n)$ by using $O(n^2)$ $\Pi_4$ gates. This circuit leads to a new (to our knowledge) factorization of $P_{2^n}$ in terms of $\Pi_4$ as 
\begin{equation}
P_{2^n} = ((\underbrace{\Pi_4 \otimes \Pi_4 \cdots \otimes \Pi_4}_{\frac{n}{2}})
(I_2 \otimes \underbrace{\Pi_4 \otimes \cdots \otimes \Pi_4}_{\frac{n}{2} -1} \otimes I_2))^{\frac{n}{2}} 
\end{equation}
for $n$ even, and 
\begin{equation}
P_{2^n} = ((I_2 \otimes \underbrace{\Pi_4 \otimes \cdots \otimes \Pi_4}_{\frac{n-1}{2}})
(\underbrace{\Pi_4  \otimes \cdots \, \Pi_4}_{\frac{n-1}{2}} \otimes I_2))^{\frac{n-1}{2}} 
(I_2 \otimes \underbrace{\Pi_4 \otimes  \cdots \otimes \Pi_4}_{\frac{n-1}{2}})
\end{equation}
for $n$ odd.
 
It should be emphasized that this new factorization of $P_{2^n}$ is less efficient than other schemes, e.g., the use of (5) for a {\it classical implementation} (see also [16] for further discussion). However, this factorization is more efficient for a {\it quantum implementation} of $P_{2^n}$. In fact, a quantum implementation of $P_{2^n}$ by using (5) and (17) will result in a complexity of $O(n^2)$ by using $O(n^2)$ $\Pi_4$ gates.     

As will be shown, the development of {\it complete} and efficient circuits for implementation of wavelet transforms requires a mechanism for implementation of conditional operators of the forms $\Pi_{2^i} \oplus I_{2^n - 2^i}$ and $P_{2^i} \oplus I_{2^n - 2^i}$, for some $i$. The key enabling factor for a successful implementation of such conditional operators is the use of factorizations similar to (17) and (19)-(20) or, alternatively, circuits similar to those in Figures 3 and 4, along with the conditional operators involving $\Pi_4$ gates.     

\section{Quantum Wavelet Algorithms}

\subsection{Wavelet Pyramidal and Packet Algorithms}

Given a wavelet kernel, its corresponding wavelet transform is usually performed according to a packet algorithm (PAA) or a pyramid algorithm (PYA). The first step in devising quantum counterparts of these algorithms is the development of suitable factorizations. Consider the Daubechies fourth-order wavelet kernel of dimension $2^i$, denoted as $D^{(4)}_{2^i}$. First level factorizations of PAA and PYA for a $2^n$-dimensional vector are given as  
\begin{equation}
PAA = (I_{2^{n-2}} \otimes D^{(4)}_4)(I_{2^{n-3}} \otimes \Pi_8) \cdots  
(I_{2^{n-i}} \otimes D^{(4)}_{2^i})(I_{2^{n-i-1}} \otimes \Pi_{2^{i+1}}) \cdots (I_2 \otimes D^{(4)}_{2^{n-1}}) \Pi_{2^n}D^{(4)}_{2^n}
\end{equation}
\begin{equation}
PYA = (D^{(4)}_4 \oplus I_{2^n-4})(\Pi_8 \oplus I_{2^n-8}) \cdots  
(D^{(4)}_{2^i} \oplus I_{2^n - 2^i})(\Pi_{2^{i+1}} \oplus I_{2^n - 2^{i+1}}) \cdots 
\Pi_{2^n}D^{(4)}_{2^n}
\end{equation}
These factorizations allow a first level analysis of the feasibility and efficiency of quantum implementations of the packet and pyramid algorithms. To see this, suppose we have a practically realizable and efficient, i.e., $O(i)$, quantum algorithm for implementation of $D^{(4)}_{2^i}$. For the packet algorithm, the operators $(I_{2^{n-i}} \otimes D^{(4)}_{2^i})$ can be directly and efficiently implemented by using the algorithm for $D^{(4)}_{2^i}$. Also, using the factorization of $\Pi_{2^i}$, given by (17), the operators $(I_{2^{n-i}} \otimes \Pi_{2^i})$ can be implemented efficiently in $O(i)$. 

For the pyramid algorithm, the existence of an algorithm for $D^{(4)}_{2^i}$ does not automatically imply an efficient algorithm for implementation of the conditional operators $(D^{(4)}_{2^i} \oplus I_{2^n - 2^i})$. An example of such a case is discussed in Sec. 4.4. Thus, careful analysis is needed to establish both the feasibility and efficiency of implementation of the conditional operators $(D^{(4)}_{2^i} \oplus I_{2^n - 2^i})$ by using the algorithm for $D^{(4)}_{2^i}$. Note, however, that the conditional operators $(\Pi_{2^i} \oplus I_{2^n - 2^i})$ can be efficiently implemented in $O(i)$ by using the factorization in (17) and the conditional $\Pi_4$ gates. 

The above analysis can be extended to any wavelet kernel (WK) and summarized as follows:

\begin{itemize}
\item Packet algorithm: A physically realizable and efficient algorithm for the WK along with the use of (17) leads to a physically realizable and efficient implementation of the packet algorithm.

\item Pyramid algorithm: A physically realizable and efficient algorithm for the WK does not automatically lead to an
implementation of the conditional operators involving WK (and hence the pyramid algorithm) but the conditional operators $(\Pi_{2^i} \oplus I_{2^n - 2^i})$ can be efficiently implemented by using the factorization in (17) and the conditional $\Pi_4$ gates.   

\end{itemize}

\subsection{Haar Wavelet Factorization and Implementation}

The Haar transform can be defined from the Haar functions [17]. Hoyer [20] used a recursive definition of Haar matrices based on the {\it generalized Kronecker product} (see also [17] for similar definitions) and developed a factorization of $H_{2^n}$ as  
\begin{eqnarray}
H_{2^n}  = & (I_{2^{n-1}} \otimes W) \cdots (I_{2^{n-i}} \otimes W \oplus 
I_{2^n - 2^{n-i+1}}) \, \cdots \, (W \oplus I_{2^n - 2}) \times \nonumber \\ 
& (\Pi_4 \oplus I_{2^n - 4}) \, \cdots \, (\Pi_{2^i} \oplus I_{2^n - 2^i}) \, \cdots \, (\Pi_{2^{n-1}} \oplus I_{2^{n-1}}) \Pi_{2^n} 
\end{eqnarray} 
Hoyer's circuit for implementation of (23) is shown in Fig 5. However, this represents an {\it incomplete} solution for   
quantum implementation and subsequent complexity analysis of the Haar transform. To see this, let 
\begin{equation}
H^{(1)}_{2^n} = (I_{2^{n-1}} \otimes W) \cdots (I_{2^{n-i}} \otimes W \oplus 
I_{2^n - 2^{n-i+1}}) \, \cdots \, (W \oplus I_{2^n - 2})
\end{equation}
\begin{equation}
H^{(2)}_{2^n} = (\Pi_4 \oplus I_{2^n - 4}) \, \cdots \, (\Pi_{2^i} \oplus I_{2^n - 2^i}) \, \cdots \, (\Pi_{2^{n-1}} \oplus I_{2^{n-1}}) \Pi_{2^n} 
\end{equation}
Clearly, the operator $H^{(1)}_{2^n}$ can be implemented in $O(n)$ by using $O(n)$ conditional $W$ gates. But the feasibility of practical implementation of the operator $H^{(2)}_{2^n}$ and its complexity (and consequently those of the factorization in (23)) cannot be assessed unless a mechanism for implementation of the terms $(\Pi_{2^i} \oplus I_{2^n - 2^i})$ is devised. 

However, by using the factorizations and circuits similar to (17) and Figure 3, it can be easily shown that the operators  $(\Pi_{2^i} \oplus I_{2^n - 2^i})$ can be implemented in $O(i)$ by using $O(i)$ conditional $\Pi_4$ gates (or, Controlled$^k$-NOT gates). This leads to the implementation of $H^{(2)}_{2^n}$ and consequently $H_{2^n}$ in $O(n^2)$ by using $O(n^2)$ gates. This represents not only the first practically feasible quantum circuit for implementation of $H_{2^n}$ but also the first complete analysis of complexity of its time and space (gates) quantum implementation. Note that, both operators $(I_{2^{n-i}} \otimes H_{2^i})$ and $(H_{2^i} \oplus I_{2^n - 2^i})$ can be directly and efficiently implemented by using the above algorithm and circuit for implementation of $H_{2^i}$. This implies both the feasibility and efficiency of the quantum implementation of the packet and pyramid algorithms by using our factorization for Haar wavelet kernel.      

\subsection{Daubechies $D^{(4)}$ Wavelet and Hoyer's Factorization}

The Daubechies fourth-order wavelet kernel of dimension $2^n$ is given in a matrix form as [22]   
\begin{equation}
D^{(4)}_{2^n} = 
\left( \begin{array}{ccccccccccc} 
c_0 & c_1 & c_2 & c_3 \\
c_3 & -c_2 & c_1 & -c_0 \\
 &  & c_0 & c_1 & c_2 & c_3 \\
 &  & c_3 & -c_2 & c_1 & -c_0 \\ 
\vdots & \vdots & & & & & \ddots \\
 &  &  &  &  &  &  & c_0 & c_1 & c_2 & c_3 \\
 &  &  &  &  &  &  & c_3 & -c_2 & c_1 & -c_0 \\
c_2 & c_3 &  &  &  &  &  &  &  & c_0 & c_1 \\ 
c_1 & -c_0 &  &  &  &  &  &  &  & c_3 & -c_2  
\end{array} \right)  
\end{equation}
where $c_0 = \frac {(1 + \sqrt {3})} {4 \sqrt {2}}$, $c_1 = \frac {(3 + \sqrt {3})} {4 \sqrt {2}}$, $c_2 = \frac {(3 - \sqrt {3})} {4 \sqrt {2}}$, and $c_3 = \frac {(1 - \sqrt {3})} {4 \sqrt {2}}$. For classical computation and given its sparse structure, the application of $D^{(4)}_{2^n}$ can be performed with an optimal cost of $O(2^n)$. However, the matrix $D^{(4)}_{2^n}$, as given by (26), is not suitable for a quantum implementation. To achieve a feasible and efficient quantum implementation, a suitable factorization of $D^{(4)}_{2^n}$ needs to be developed. Hoyer [20] proposed a factorization of $D^{(4)}_{2^n}$ as    
\begin{equation}
D^{(4)}_{2^n} = (I_{2^{n-1}} \otimes C_1) S_{2^n}(I_{2^{n-1}} \otimes C_0)
\end{equation}
where 
\begin{equation}
C_0 =  
2 \left( \begin{array}{cc}
c_4 & -c_2 \\
-c_2 & c_4 
\end{array} \right)  
\mbox{ and } 
C_1 = \frac{1}{2} 
\left( \begin{array}{cc}
\frac{c_0}{c_4} & 1 \\
1 & \frac{c_1}{c_2} 
\end{array} \right) 
\end{equation}
and $S_{2^n}$ is a permutation matrix with a classical description given by 
\begin{equation}
S_{ij} = \left\{ \begin{array}{cc}
1 & \mbox{  if $i = j$ and $i$ is even, or if $i+2 = j$ \, (mod $2^n$)} \\
0 & \mbox{  otherwise}
\end{array} \right.
\end{equation}
Hoyer's block-level circuit for implementation of (27) is shown in Figure 6. Clearly, the main issue for a practical quantum implementation and subsequent complexity analysis of (27) is the quantum implementation of matrix $S_{2^n}$. To this end,  
Hoyer discovered a quantum arithmetic description of $S_{2^n}$ as      
\begin{equation}
S_{2^n}: \, \vert a_{n-1} \, a_{n-2} \, \cdots a_1 \, a_0 \rangle \, 
\longmapsto \vert b_{n-1} \, b_{n-2} \, \cdots b_1 \, b_0   \rangle
\end{equation}
where 
\begin{equation}
b_i = \left\{ \begin{array}{cc}
a_i - 2  \mbox{ \, (mod $n$)}, & \mbox{if $i$ is odd} \\
a_i & \mbox{otherwise}
\end{array} \right.
\end{equation}   
As suggested by Hoyer, this description of $S_{2^n}$ then allows its quantum implementation by using quantum arithmetic circuits of [18] with a complexity of $O(n)$. This algorithm can be directly extended for implementation of the operators 
$(I_{2^{n-i}} \otimes D^{(4)}_{2^i})$ and hence the packet algorithm. However, the feasibility and efficiency of an  implementation of the operators $(I_{2^{n-i}} \oplus D^{(4)}_{2^i})$ and thus the pyramid algorithm needs further analysis.
  
\section{Fast Quantum Algorithms and Circuits for Implementation of Daubechies $D^{(4)}$ Wavelet}

In this section, we develop a new factorization of the Daubechies $D^{(4)}$ wavelet. This factorization leads to three new and efficient circuits, including one using the circuit for QFT, for implementation of Daubechies $D^{(4)}$ wavelet.   
  
\subsection{A New Factorization of Daubechies $D^{(4)}$ Wavelet}

We develop a new factorization of the Daubechies $D^{(4)}$ wavelet transform by showing that the permutation matrix $S_{2^n}$ can be written as a product of two permutation matrices as 
\begin{equation}
S_{2^n} = Q_{2^n}R_{2^n}
\end{equation}
where $Q_{2^n}$ is the {\it downshift permutation matrix} [16] given by 
\begin{equation}
Q_{2^n} = \left( \begin{array}{ccccccc}
0 & 1 \\
0 & 0 & 1 \\
0 & 0 & 0 & 1 \\
\vdots & \vdots & \vdots & & \ddots \\
0 & 0 & \cdots & 0 & 0 & 1 \cr
1 & 0 & \cdots & 0 & 0 & 0 
\end{array} \right)  
\end{equation}
and $R_{2^n}$ is a permutation matrix given by  
\begin{equation}
R_{2^n} = \left( \begin{array}{cccccccc}
0 & 1 & 0 & 0 & 0 \\
1 & 0 & 0 & 0 & 0 \\
0 & 0 & 0 & 1 & 0 \\
0 & 0 & 1 & 0 & 0 \\
 & \ddots & \ddots & \ddots & \ddots \\
 & & & & & & 0 & 1 \cr
 & & & & & & 1 & 0 
\end{array} \right)  
\end{equation}
The matrix $R_{2^n}$ can be written as 
\begin{equation}
R_{2^n} = I_{2^{n-1}} \otimes N   
\end{equation}
where $
N = 
\left( \begin{array}{cc} 
0 & 1 \\
1 & 0  
\end{array} \right)
$. 
Substituting (35) and (32) into (27), a new factorization of $D^{(4)}_{2^n}$ is derived as 
\begin{equation}
D^{(4)}_{2^n} = (I_{2^{n-1}} \otimes C_1) Q_{2^n}(I_{2^{n-1}} \otimes N)
(I_{2^{n-1}} \otimes C_0) = (I_{2^{n-1}} \otimes C_1) Q_{2^n}
(I_{2^{n-1}} \otimes C_0^\prime)
\end{equation}
where 
\begin{equation}
C_0^\prime = N.C_0 = 2 
\left( \begin{array}{cc} 
-c_2 & c_4 \\
c_4 & -c_2 
\end{array} \right)
\end{equation}
Fig. 7 shows a block-level implementation of (36). Clearly, the main issue for a practical quantum gate-level implementation and subsequent complexity analysis of (36) is the quantum implementation of matrix $Q_{2^n}$. In the following, we present three circuits for quantum implementation of matrix $Q_{2^n}$. 
  
\subsection{Quantum Arithmetic Implementation of Permutation Matrix $Q_{2^n}$}

A first circuit for implementation of matrix $Q_{2^n}$ is developed based on its description as a {\it quantum arithmetic operator}. We have discovered such a quantum arithmetic description of $Q_{2^n}$ as     
\begin{equation}
Q_{2^n}: \, \vert a_{n-1} \, a_{n-2} \, \cdots a_1 \, a_0 \rangle \, 
\longmapsto \vert b_{n-1} \, b_{n-2} \, \cdots b_1 \, b_0   \rangle
\end{equation}
where 
\begin{equation}
b_i = a_i - 1  \mbox { \, (mod $n$)}
\end{equation}
This description of $Q_{2^n}$ allows its quantum implementation by using quantum arithmetic circuit of [18] with a complexity of $O(n)$. Note, however, that the arithmetic description of $Q_{2^n}$ is simpler than that of $S_{2^n}$ since it does not involve conditional quantum arithmetic operations (i.e., the same operation is applied to all qubits). This algorithm for quantum implementation of $Q_{2^n}$ and hence $D^{(4)}_{2^n}$ can be directly extended for implementation of the operators $(I_{2^{n-i}} \otimes D^{(4)}_{2^i})$ and hence the packet algorithm. However, the feasibility and efficiency of an implementation of the operators $(I_{2^{n-i}} \oplus D^{(4)}_{2^i})$ and thus the pyramid algorithm needs further analysis.

\subsection{Quantum FFT Factorization of Permutation Matrix $Q_{2^n}$} 
 
A direct and efficient factorization and subsequent circuit for implementation of $Q_{2^n}$ (and hence Daubechies $D^{(4)}$ wavelet) can be derived by using the FFT algorithm. This factorization is based on the observation that $Q_{2^n}$ can be described in terms of FFT as [16]
\begin{equation}
Q_{2^n} = F_{2^n} T_{2^n} F^*_{2^n}
\end{equation}
where $T_{2^n}$ is a diagonal matrix given as 
$T_{2^n} = \mbox {Diag} \{1, \, \omega_{2^n}, \, \omega_{2^n}^2, \, \ldots , 
\omega_{2^n}^{2^n -1} \}$ with  
$\omega_{2^n} = e^{{-2 \iota \pi} \over {2^n}}$ (* indicates conjugate transpose). As will be seen, it is more efficient to use the Cooley-Tukey factorization, given by (7), and write (40) as  
\begin{equation}
Q_{2^n} = {\underline F}_{2^n} P_{2^n} T_{2^n}P_{2^n}{\underline F}^*_{2^n}  
\end{equation}
It can be shown that the matrix $T_{2^n}$ has a factorization as 
\begin{equation}
T_{2^n} = (G(\omega_{2^n}^{2^{n-1}}) \otimes I_{2^{n-1}}) \cdots 
(I_{2^{i-1}} \otimes G(\omega_{2^n}^{2^{n-i}}) \otimes I_{2^{n-i}}) \cdots 
(I_{2^{n-1}} \otimes G(\omega_{2^n}))
\end{equation}
where $G(\omega_{2^n}^k) = \mbox {Diag} \{1, \, \omega_{2^n}^k \} = 
\left( \begin{array}{cc}
1 & 0 \\
0 & \omega_{2^n}^k 
\end{array} \right)
$. This factorization leads to an efficient implementation of $T_{2^n}$ by using $n$ single qubit $G(\omega_{2^n}^k)$ gates as shown in Fig. 8. Together with the circuit for implementation of $P_{2^n}$ (Fig. 4) and the circuit for implementation of FFT (Fig. 1), they represent a complete gate-level implementation of $D^{(4)}_{2^n}$. 

However, a more efficient circuit can be derived by avoiding the explicit implementation of $P_{2^n}$ by showing that the operator 
\begin{equation}
P_{2^n}T_{2^n}P_{2^n} = P_{2^n}(G(\omega_{2^n}^{2^{n-1}}) \otimes I_{2^{n-1}})\cdots 
(I_{2^{i-1}} \otimes G(\omega_{2^n}^{2^{n-i}}) \otimes I_{2^{n-i}}) \cdots 
(I_{2^{n-1}} \otimes G(\omega_{2^n}))P_{2^n}
\end{equation} 
can be efficiently implemented by simply reversing the order of gates in Fig. 8. This is established by the following lemma:
\vspace{0.1in}

\noindent {\bf Lemma 1.} 
\vspace{0.1in} 
\begin{equation}
P_{2^n}(G(\omega_{2^n}^{2^{n-1}}) \otimes I_{2^{n-1}}) = 
(I_{2^{n-1}} \otimes G(\omega_{2^n}^{2^{n-1}}))P_{2^n}
\end{equation}
\begin{equation}
P_{2^n} (I_{2^{n-j}} \otimes G(\omega_{2^n}^{2^{j-1}}) \otimes I_{2^{j-1}}) = 
(I_{2^{j-1}} \otimes G(\omega_{2^n}^{2^{j-1}}) \otimes I_{2^{n-j}})P_{2^n}
\end{equation}
\begin{equation}
P_{2^n}(I_{2^{n-1}} \otimes G(\omega_{2^n})) = (G(\omega_{2^n}) \otimes 
I_{2^{n-1}})P_{2^n}
\end{equation}
\noindent {\bf\it Proof.} This lemma can be easily proved based on the physical interpretation of operations in (44)-(46). The left-hand side of (44) implies first an operation, i.e., application of $G(\omega_{2^n}^{2^{n-1}})$, on the last qubit and then application of $P_{2^n}$ on all the qubits, i.e., reversing the order of qubits. However, this is equivalent to first reversing the order of qubits, i.e., applying $P_{2^n}$, and then applying $G(\omega_{2^n}^{2^{n-1}})$, on the first qubit which is the operation described by the right-hand side of (44). Similarly, the left-hand side of (45) implies first application of $G(\omega_{2^n}^{2^{i-1}})$ on the $(n-i)$th qubit and then reversing the order of qubits. This is equivalent to first reversing the order of qubits and then applying $G(\omega_{2^n}^{2^{i-1}})$ on the $i$th qubit which is the operations described by the right hand side of (45). In a same fashion, the left hand side of (46) implies first application of $G(\omega_{2^n})$ on the first qubit and then reversing the order of qubits which is equivalent to first reversing the order of qubits and then applying $G(\omega_{2^n}^{2^{n-1}})$ on the last qubit, that is, the operations in right-hand side of (46).

Applying (44)-(46) to (43) from left to right and noting that, due to the symmetry of $P_{2^n}$, we have 
$P_{2^n}P_{2^n} = I_{2^n}$, it then follows that 
\begin{equation}
P_{2^n}T_{2^n}P_{2^n} = (I_{2^{n-1}} \otimes G(\omega_{2^n}^{2^{n-1}})) \cdots 
(I_{2^{n-i}} \otimes G(\omega_{2^n}^{2^{n-i}}) \otimes I_{2^{i-1}}) \cdots
(G(\omega_{2^n}) \otimes I_{2^{n-1}})
\end{equation}
The circuit for implementation of (47) is shown in Fig.9 which, as can be seen, has been obtained by reversing the order of gates in Fig. 8. Note that, the use of (47), which is a direct consequence of using the Cooley-Tukey factorization, enables the implementation of (40) without explicit implementation of $ P_{2^n}$. 

Using (40) and (47), the complexity of the implementation of $Q_{2^n}$ and thus $D^{(4)}_{2^n}$ is the same as of the quantum FFT, that is, $O(n^2)$ for an exact implementation and $O(nm)$ for an approximation of order $m$ [15]. Note 
that, by using (47), (40), and (36) both operators $(I_{2^{n-i}} \otimes D^{(4)}_{2^i})$ and 
$(D^{(4)}_{2^i} \oplus I_{2^n - 2^i})$ can be directly implemented. This implies both the feasibility and efficiency 
of the quantum implementation of the packet and pyramid algorithms by using this algorithm for quantum implementation of $D^{(4)}_{2^n}$.      
	
\subsection{A Direct Recursive Factorization of Permutation Matrix $Q_{2^n}$} 

A new direct and recursive factorization of $Q_{2^n}$ can be derived based on a similarity transformation of $Q_{2^n}$ by using $\Pi_{2^n}$ as    
\begin{equation}
\Pi^t_{2^n}Q_{2^n}\Pi_{2^n} = 
\left( \begin{array}{cc}
0 & I_{2^{n-1}} \\
Q_{2^{n-1}} & 0 
\end{array} \right)
\end{equation} 
which can be written as 
\begin{equation}
\Pi^t_{2^n}Q_{2^n}\Pi_{2^n} = 
\left( \begin{array}{cc}
0 & I_{2^{n-1}} \\
I_{2^{n-1}} & 0 
\end{array} \right) 
\left( \begin{array}{cc}
Q_{2^{n-1}} & 0 \\
0 & I_{2^{n-1}} 
\end{array} \right) = 
(N \otimes I_{2^{n-1}})(Q_{2^{n-1}} \oplus I_{2^{n-1}})
\end{equation} 
from which $Q_{2^n}$ can be calculated as 
\begin{equation} 
Q_{2^n} = \Pi_{2^n}(N \otimes I_{2^{n-1}})(Q_{2^{n-1}} \oplus 
I_{2^{n-1}})\Pi^t_{2^n}  
\end{equation}
Replacing a similar factorization of $Q_{2^{n-1}}$ into (50), we get 
\begin{equation}
Q_{2^n} = \Pi_{2^n}(N \otimes I_{2^{n-1}})
(\Pi_{2^{n-1}}(N \otimes I_{2^{n-2}})(Q_{2^{n-2}} \oplus I_{2^{n-2}})  \Pi^t_{2^{n-1}} \oplus I_{2^{n-1}})\Pi^t_{2^n}
\end{equation}
By using the identity 
\begin{equation}
\Pi_{2^{n-1}} A \Pi^t_{2^{n-1}} \oplus I_{2^{n-1}} = 
(I_2 \otimes \Pi_{2^{n-1}})(A \oplus I_{2^{n-1}})(I_2 \otimes \Pi^t_{2^{n-1}})
\end{equation}
for any matrix $A \varepsilon \Re^{2^{n-1} \times 2^{n-1}}$, (51) can be then  written as
\begin{equation}
Q_{2^n} = \Pi_{2^n}(N \otimes I_{2^{n-1}})(I_2 \otimes \Pi_{2^{n-1}})
((N \otimes I_{2^{n-2}})(Q_{2^{n-2}} \oplus I_{2^{n-2}})\oplus I_{2^{n-1}})  
(I_2 \otimes \Pi^t_{2^{n-1}})\Pi^t_{2^n} 
\end{equation}
Using the identity   
\begin{eqnarray}
(N \otimes I_{2^{n-2}})(Q_{2^{n-2}} \oplus I_{2^{n-2}})\oplus I_{2^{n-1}} & = & 
(N \otimes I_{2^{n-2}} \oplus I_{2^{n-1}})(Q_{2^{n-2}} \oplus
I_{2^{n-2}} \oplus I_{2^{n-1}})  \nonumber \\ 
& = & (N \otimes I_{2^{n-2}} \oplus I_{2^{n-1}})(Q_{2^{n-2}} \oplus I_{3.2^{n-2}})
\end{eqnarray}
(53) is now written as  
\begin{equation}
Q_{2^n} = \Pi_{2^n}(N \otimes I_{2^{n-1}})(I_2 \otimes \Pi_{2^{n-1}})
(N \otimes I_{2^{n-2}} \oplus I_{2^{n-1}})(Q_{2^{n-2}} \oplus I_{2^n - 2^{n-2}})
(I_2 \otimes \Pi^t_{2^{n-1}})\Pi^t_{2^n}
\end{equation}
Repeating the same procedures for all $Q_{2^i}$, for $i = n-3$ to 1, and noting that $Q_2 = N$, it then follows 
\begin{eqnarray}
Q_{2^n} & = \Pi_{2^n}(N \otimes I_{2^{n-1}})(I_2 \otimes \Pi_{2^{n-1}})
(N \otimes I_{2^{n-2}} \oplus I_{2^{n-1}})(I_4 \otimes \Pi_{2^{n-2}})
(N \otimes I_{2^{n-3}} \oplus I_{2^n - 2^{n-2}}) \cdots \nonumber \\
 & (I_{2^{n-2}} \otimes \Pi_4)(N \otimes I_2 \oplus I_{2^n -4})
(N \oplus I_{2^n -2})(I_{2^{n-2}} \otimes \Pi^t_4) \cdots 
(I_2 \otimes \Pi^t_{2^{n-1}})\Pi^t_{2^n}
\end{eqnarray}

The above expression of $Q_{2^n}$ can be further simplified by exploiting the fact that (see Appendix for the proof) every operator of the form $(I_{2^i} \otimes \Pi_{2^{n-i}})$, for $i = n-2$ to $1$, commutes with all operators of the form $(N \otimes I_{2^{n-j}} \oplus I_{2^n - 2^{n-j+1}})$, for $j = i$ to $1$. Using this commutative property, (56) can be now written as 
\begin{eqnarray}
Q_{2^n} & = \Pi_{2^n}(I_2 \otimes \Pi_{2^{n-1}})(I_4 \otimes
\Pi_{2^{n-2}}) \cdots (I_{2^{n-2}} \otimes \Pi_{4}) (N \otimes
I_{2^{n-1}}) (N \otimes I_{2^{n-2}} \oplus I_{2^{n-1}}) \cdots  \nonumber \\
 & \qquad (N \otimes I_2 \oplus I_{2^n -4}) (N \oplus I_{2^n
-2})(I_{2^{n-2}} \otimes \Pi^t_4) \cdots (I_2 \otimes
\Pi^t_{2^{n-1}})\Pi^t_{2^n} 
\end{eqnarray}
Using the factorization of $P_{2^n}$ given in (5), we then have  
\begin{equation}
Q_{2^n} = P_{2^n}(N \otimes I_{2^{n-1}})(N \otimes I_{2^{n-2}} \oplus I_{2^{n-1}}) \cdots (N \otimes I_2 \oplus I_{2^n -4})(N \oplus I_{2^n -2})P_{2^n}
\end{equation}
Substituting (58) into (36), a factorization of $D^{(4)}_{2^n}$ is then obtained as
\begin{equation}
D^{(4)}_{2^n} = (I_{2^{n-1}} \otimes C_1) P_{2^n}(N \otimes I_{2^{n-1}})(N \otimes I_{2^{n-2}} \oplus I_{2^{n-1}}) \cdots (N \otimes I_2 \oplus I_{2^n -4})(N \oplus I_{2^n -2})P_{2^n} (I_{2^{n-1}} \otimes C_0^\prime)
\end{equation}
Using Lemma 1, it then follows that 
\begin{equation}
D^{(4)}_{2^n} = P_{2^n}(C_1 \otimes I_{2^{n-1}})(N \otimes I_{2^{n-1}})(N \otimes I_{2^{n-2}} \oplus I_{2^{n-1}}) \cdots (N \otimes I_2 \oplus I_{2^n -4})(N \oplus I_{2^n -2})(C_0^\prime \otimes I_{2^{n-1}})P_{2^n} 
\end{equation}
A circuit for implementation of $D^{(4)}_{2^n}$, based on (60), is shown in Fig. 10. Together with the circuit for  implementation of $P_{2^n}$, shown in Fig. 4, they represent a complete gate-level circuit for implementation of $D^{(4)}_{2^n}$ with an optimal complexity of $O(n)$.  

Using (60) and (19)-(20), the operators $(I_{2^{n-i}} \otimes D^{(4)}_{2^i})$ can be directly and efficiently implemented with a complexity of $O(i)$. This implies both the feasibility and efficiency of the implementation of the packet algorithm by using this algorithm for $D^{(4)}_{2^n}$ wavelet kernel. However, this algorithm is less efficient for implementation of the operators $(D^{(4)}_{2^i} \oplus I_{2^n - 2^i})$ and hence the pyramid algorithm. To see this, note that, the implementation of the operators $(D^{(4)}_{2^i} \oplus I_{2^n - 2^i})$, by using (60), requires the implementation of the conditional operators $(P_{2^i} \oplus I_{2^n - 2^i})$. However, these conditional operators cannot be directly implemented by using (19) and (20). An alternative solution is to use the factorization of $P_{2^i}$ in (5) and the conditional operators $(\Pi_{2^i} \oplus I_{2^n - 2^i})$. However, this leads to a complexity of $O(i^2)$ for implementation of operators $(P_{2^i} \oplus I_{2^n - 2^i})$ and hence the operators $(D^{(4)}_{2^i} \oplus I_{2^n - 2^i})$. Therefore, while (60) is optimal for implementation of $D^{(4)}_{2^i}$ and the packet algorithm, it is not efficient for implementation of the pyramid algorithm.

It should be emphasized that this recursive factorization of $Q_{2^n}$, originated by the similarity transformation in (48) and given by (56) and (58), was not previously known in classical computing. Note that, the permutation matrices $\Pi_{2^n}$ and, particularly, $P_{2^n}$ are much harder (in terms of data movement pattern) for a classical implementation than $Q_{2^n}$. In this sense, such a factorization of $Q_{2^n}$ is rather counterintuitive from a classical computing point of view since it involves the use of permutation matrices $\Pi_{2^n}$ and $P_{2^n}$ and thus it is highly inefficient for a classical implementation.        
 
\section{Discussion and Conclusion}

In this paper, we developed fast algorithms and efficient circuits for quantum wavelet transforms. Assuming an efficient quantum circuit for a given wavelet kernel and starting with a high level description of the packet and pyramid algorithms, we analyzed the feasibility and efficiency of the implementation of the packet and pyramid algorithms by using the given wavelet kernel. We also developed efficient and complete gate-level circuits for two representative wavelet kernels, the Haar and Daubechies $D^{(4)}$ kernels. We gave the first complete time and space complexity analysis of the quantum Haar wavelet transform. We also described three complete circuits for Daubechies $D^{(4)}$ wavelet kernel. In particular, we showed that Daubechies $D^{(4)}$ kernel can be implemented by using the circuit for QFT. Given the problem of decoherence, exploitation of parallelism in quantum computation is a key issue in practical implementation of a given computation. To this end, we are currently analyzing the algorithms of this paper in terms of their parallel efficiency and developing more efficient parallel quantum wavelet algorithms. 

As shown in this paper, permutation matrices play a pivotal role in the development of quantum wavelet transforms. In fact, not only they arise explicitly in the packet and pyramid algorithms but also they play a key role in factorization of wavelet kernels. For classical computing, the implementation of permutation matrices is trivial. However, for quantum computing, it represents a challenging task and demands new, unconventional, and even counterintuitive (from a classical computing view point) techniques. For example, note that most of the factorizations developed in paper for permutation matrices $\Pi_{2^n}$, $P_{2^n}$, and $Q_{2^n}$ were not previously known in classical computing and, in fact, they are not at all efficient for a classical implementation. Also, implementation of the permutation matrices reveals some of the surprises of quantum computing in contrast to classical computing. In the sense that, certain operations that are hard to implement in classical computing are easier to implement in quantum computing and vice versa. As a concrete example, note that while the classical implementation of permutation matrices $\Pi_{2^n}$ and (particularly) $P_{2^n}$ is much harder (in terms of data movement pattern) than the permutation matrix $Q_{2^n}$, their quantum implementation is much easier and more straightforward than $Q_{2^n}$. 

In this paper, we focussed on the set of permutation matrices arising in the development of quantum wavelet transforms and analyzed three techniques for their quantum implementation. However, it is clear that the permutation matrices will also play a major role in deriving compact and efficient factorizations, i.e., with polynomial time and space complexity, for other unitary operators by exposing and exploiting their specific structure. Therefore, we believe strongly that a more systematic study of permutation matrices is needed in order to develop further insight into efficient techniques for their implementation in quantum circuits. Such a study might eventually lead to the discovery of new and more efficient approaches for the implementation of unitary transformations and therefore quantum computation.     
\clearpage

\noindent {\bf Acknowledgement}

The research described in this paper was performed at the Jet Propulsion Laboratory (JPL), California Institute of Technology, under contract with National Aeronautics and Space Administration (NASA). This work was supported by the NASA/JPL Center for Integrated Space Microsystems (CISM), NASA/JPL Advanced Concepts Office, and NASA/JPL Autonomy and Information Technology Management Program. 

\vspace{0.1in}

\noindent {\bf Appendix: Commutation of the Operators $I_{2^i} \otimes \Pi_{2^{n-i}}$ with $N \otimes I_{2^{n-j}} \oplus I_{2^n - 2^{n-j+1}}$}

We first prove that every operator of the form $I_{2^i} \otimes \Pi_{2^{n-i}}$, for $i = n-2$ to $1$, commutes with all the operators of the form $N \otimes I_{2^{n-j}} \oplus I_{2^n - 2^{n-j+1}}$, for $j = i$ to $2$, by simply showing  that 
\begin{equation}
(I_{2^i} \otimes \Pi_{2^{n-i}})(N \otimes I_{2^{n-j}} \oplus I_{2^n - 2^(n-j+1}) = (N \otimes I_{2^{n-j}} \oplus I_{2^n - 2^{n-j+1}})(I_{2^i} \otimes \Pi_{2^{n-i}})
\end{equation} 
The matrix $I_{2^i} \otimes \Pi_{2^{n-i}}$ is a block diagonal matrix and therefore can be written as 
\begin{equation}
I_{2^i} \otimes \Pi_{2^{n-i}} = I_2 \otimes \Pi_{2^{n-j}} \oplus I_{2^j - 2} \otimes \Pi_{2^{n-j}} 
\end{equation}       
It can be then shown that
\begin{equation}
(I_2 \otimes \Pi_{2^{n-j}} \oplus I_{2^j - 2} \otimes \Pi_{2^{n-j}})(N \otimes I_{2^{n-j}} \oplus I_{2^n - 2^{n-j+1}}) = N \otimes \Pi_{2^{n-j}} \oplus 
I_{2^j - 2} \otimes \Pi_{2^{n-j}}   
\end{equation}       
and  
\begin{equation}
(N \otimes I_{2^{n-j}} \oplus I_{2^n - 2^{n-j+1}})(I_2 \otimes \Pi_{2^{n-j}} \oplus I_{2^j - 2} \otimes \Pi_{2^{n-j}}) = N \otimes \Pi_{2^{n-j}} \oplus I_{2^j - 2} \otimes \Pi_{2^{n-j}}   
\end{equation}       
It now remains to show that every operator of the form $I_{2^i} \otimes \Pi_{2^{n-i}}$ commutes with the operator $N \otimes I_{2^{n-1}}$. This is simply proved by first using the fact that 
\begin{equation}
I_{2^i} \otimes \Pi_{2^{n-i}} = I_2 \otimes (I_{2^{i-1}} \otimes \Pi_{2^{n-i}})
\end{equation} 
and then showing that 
\begin{equation}
(I_2 \otimes (I_{2^{i-1}} \otimes \Pi_{2^{n-i}}))( N \otimes I_{2^{n-1}}) = 
(N \otimes I_{2^{n-1}})(I_2 \otimes (I_{2^{i-1}} \otimes \Pi_{2^{n-i}})) = 
N \otimes I_{2^{i-1}} \otimes \Pi_{2^{n-i}}
\end{equation}

\vspace{0.1in}




\clearpage

\input{epsf}

\begin{figure}[t]
\epsfxsize=5.0in
\centerline{\epsffile{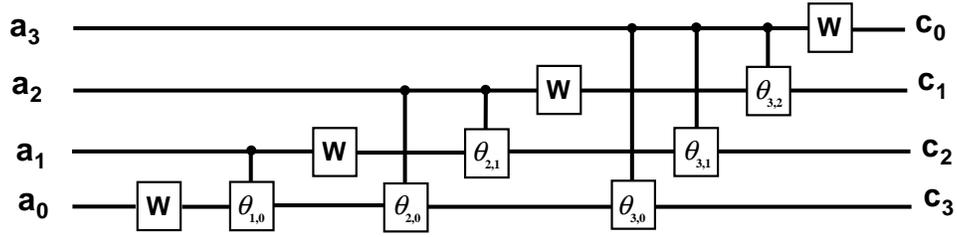}} 
\vspace{.25in}
\caption{A circuit for implementation of quantum Fourier transform, QFT (from [15]).}
\label{fig:one}
\end{figure}

\vspace{1in}

\begin{figure}[t]
\epsfxsize=5.0in
\centerline{\epsffile{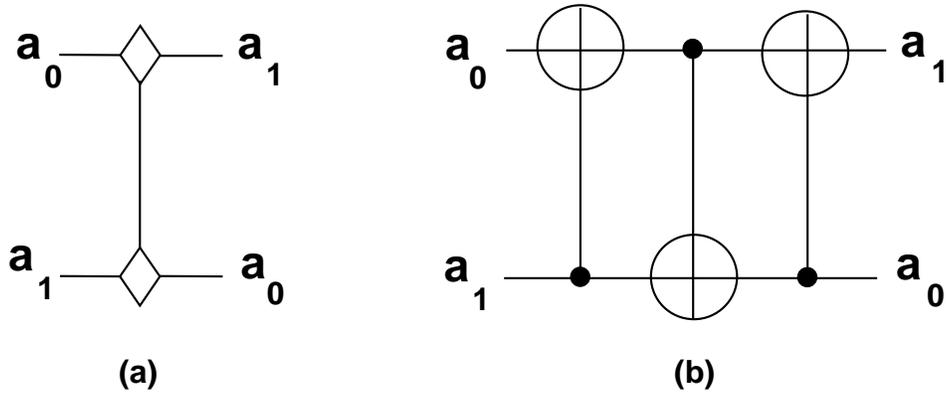}} 
\vspace{.25in}
\caption{The $\Pi_4$ gate (a) and its implementation by using three XOR (Controlled-NOT) gates (b).}
\label{fig:two}
\end{figure}
 
\vspace{1in}

\begin{figure}[t]
\epsfxsize=5.0in
\centerline{\epsffile{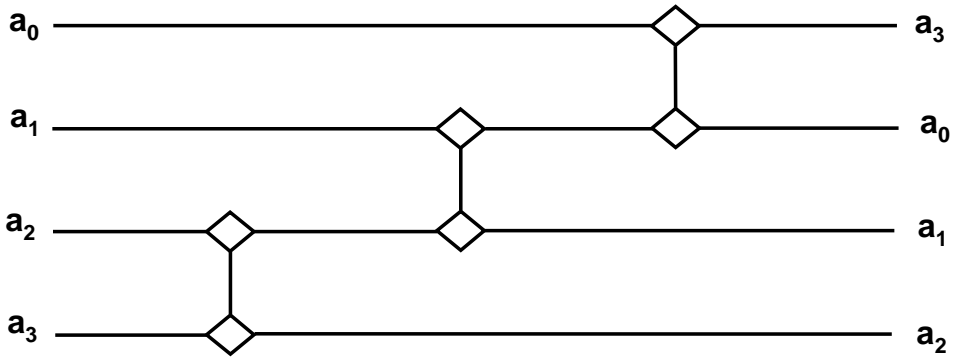}} 
\vspace{.25in}
\caption{A circuit for implementation of Perfect Shuffle permutation matrix, $\Pi_{2^n}$.}
\label{fig:three}
\end{figure}

\clearpage

\begin{figure}[t]
\epsfxsize=5.0in
\centerline{\epsffile{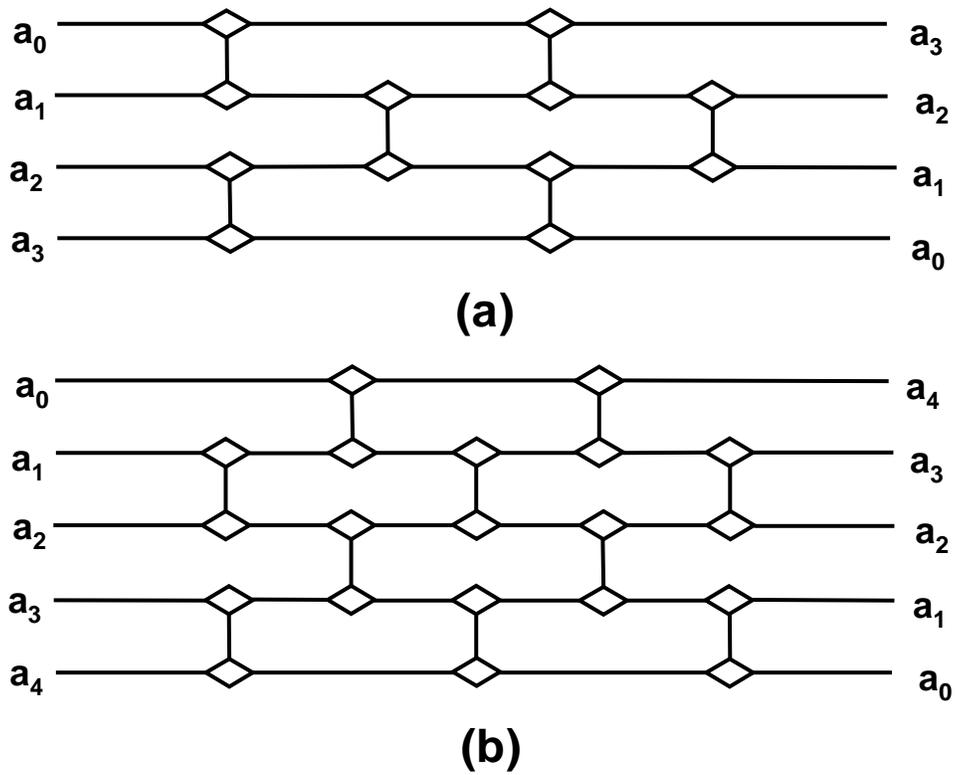}} 
\vspace{.25in}
\caption{Circuits for implementation of Bit Reversal permutation matrix, $P_{2^n}$, for $n$ even (a) and for $n$ odd (b).}
\label{fig:four}
\end{figure}

\vspace{1in}

\begin{figure}[t]
\epsfxsize=5.0in
\centerline{\epsffile{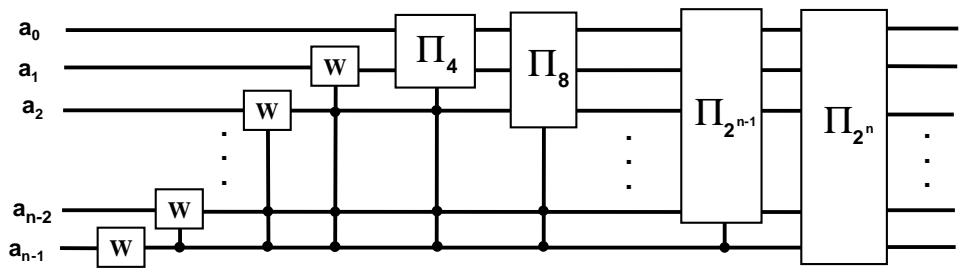}} 
\vspace{.25in}
\caption{A block-level circuit for Haar wavelet (from [20]).}
\label{fig:five}
\end{figure}

\clearpage

\begin{figure}[t]
\epsfxsize=5.0in
\centerline{\epsffile{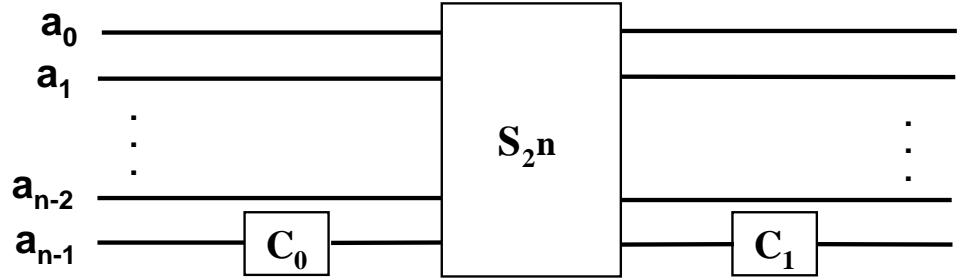}} 
\vspace{.25in}
\caption{A block-level circuit for implementation of Hoyer's factorization of $D^{(4)}_{2^n}$.}
\label{fig:six}
\end{figure}

\vspace{1in}

\begin{figure}[t]
\epsfxsize=5.0in
\centerline{\epsffile{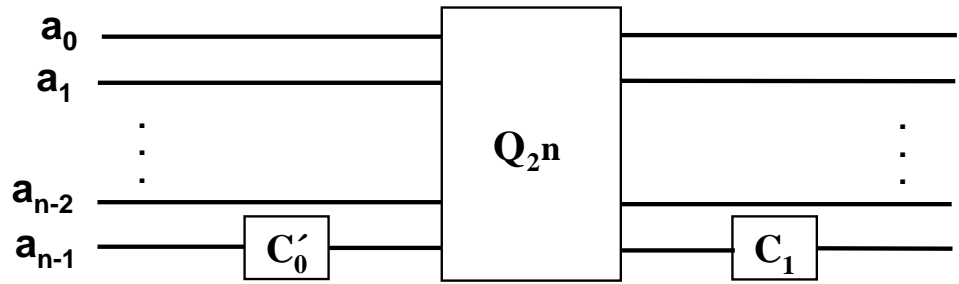}}
\vspace{.25in}
\caption{A block-level circuit for implementation of new factorization of $D^{(4)}_{2^n}$.}
\label{fig:seven}
\end{figure}

\clearpage

\begin{figure}[t]
\epsfxsize=5.0in
\centerline{\epsffile{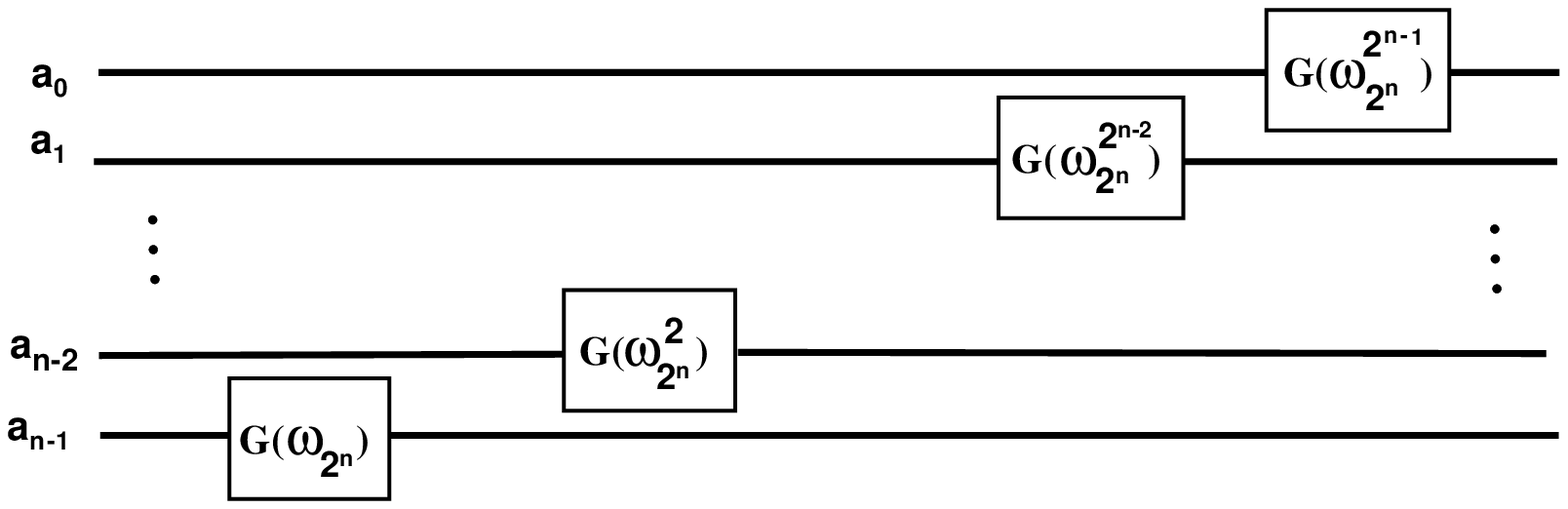}} 
\vspace{.25in}
\caption{A circuit for implementation of operator $T_{2^n}$.}
\label{fig:eight}
\end{figure}

\vspace{1in}

\begin{figure}[t]
\epsfxsize=5.0in
\centerline{\epsffile{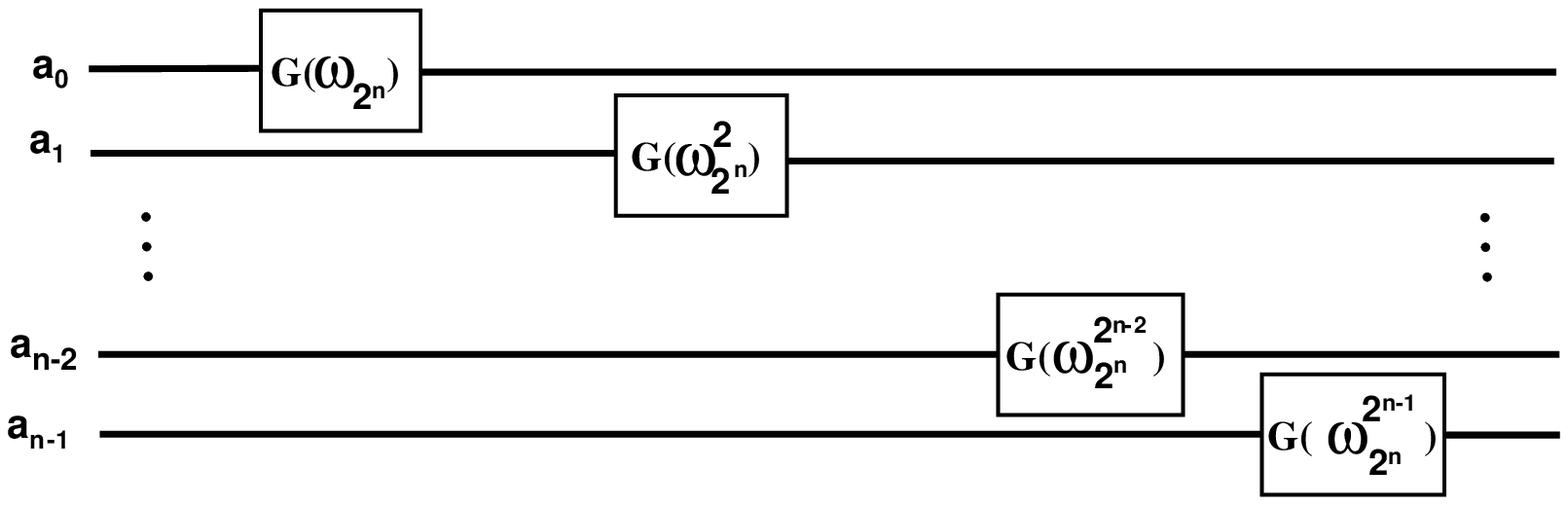}}
\vspace{.25in}
\caption{A circuit for implementation of operator $P_{2^n}T_{2^n}P_{2^n}$}
\label{fig:nine}
\end{figure}

\vspace{1in}

\begin{figure}[t]
\epsfxsize=5.0in
\centerline{\epsffile{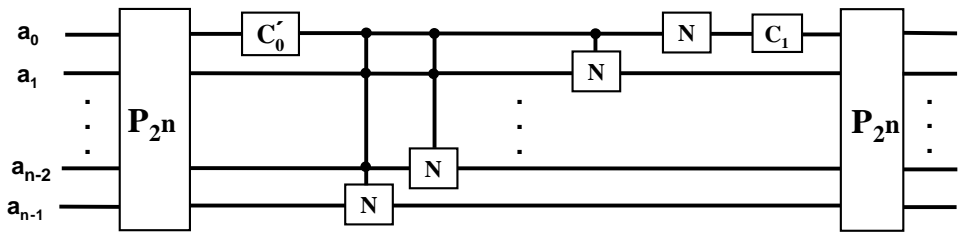}}
\vspace{.25in}
\caption{A circuit for implementation of $D^{(4)}_{2^n}$ by using recursive factorization of $Q_{2^n}$.}
\label{fig:ten}
\end{figure}

\clearpage

\end{document}